\documentclass[aps,pre,twocolumn,nofootinbib, longbibliography,preprint,10pt]{revtex4-1}

\usepackage{graphicx}
\usepackage{dcolumn}
\usepackage{bm}
\usepackage{float}
 
\usepackage[utf8]{inputenc}
\usepackage{mathtools}
\usepackage[T1]{fontenc}
\usepackage{mathptmx}
\usepackage[utf8]{inputenc}
\usepackage{graphicx}
\usepackage{subcaption}
\usepackage{amsmath}
\usepackage{graphicx}
\usepackage{subcaption}
\usepackage{amsmath}
\usepackage{gensymb}
\usepackage{amssymb}
\usepackage{dcolumn}
\usepackage{bm}
\usepackage{xr}
\usepackage{cleveref}
\usepackage{amssymb}
\usepackage{dcolumn}
\usepackage{bm}
\usepackage{xr}
\usepackage{cleveref}
\usepackage{soul}
\usepackage{color}
\usepackage{ulem}
\usepackage[usenames,dvipsnames]{xcolor}

\begin{document}

\title{A novel heterogeneous structure formed by a single multiblock copolymer chain}
\author{Artem Petrov}
\email{petrov.ai15@physics.msu.ru}
	\affiliation{Faculty of Physics, Lomonosov Moscow State University, 119991 Moscow, Russia}
	\author{Alexey Gavrilov}
	\affiliation{Faculty of Physics, Lomonosov Moscow State University, 119991 Moscow, Russia}
	\author{Alexander Chertovich}
	\affiliation{Semenov Federal Research Center for Chemical Physics, 119991 Moscow, Russia}
	\affiliation{Faculty of Physics, Lomonosov Moscow State University, 119991 Moscow, Russia}
	\date{\today}
	
\begin{abstract}
We studied structures formed by a single $(AB)_k$ multiblock copolymer chain, in which interaction between A-type beads is purely repulsive, and B-type beads tend to aggregate. We studied how attraction between A-type beads and B-type beads affects the structure of the chain. We discovered formation of an equilibrium globular structure, which had unique heterogeneous checkerboard-like distribution of contact density. Unlike the structures usually formed by a single $(AB)_k$ multiblock copolymer chain, this structure had contact enrichment at the boundaries of A and B blocks. This structure was formed by a multiblock copolymer chain, in which B-type beads could form maximum two reversible bonds either with A-type or B-type beads, A-type beads could form maximum one reversible bond with a B-type bead, and interactions between A-type beads were purely repulsive. Multiblock copolymer chains with this type of intrachain interactions can model structure of chromatin in various organisms.
\end{abstract}

\maketitle

\section{Introduction}
Melts and blends of block copolymers are extensively studied polymer systems that are widely used in the industry. The main difference between these systems and homopolymer melts is the phenomenon of microphase separation occurring in block copolymer systems. Phase diagrams of melts consisting of block copolymers with beads of two types (usually denoted as A and B) are well understood by theory \cite{leibler1980theory,semenov1985contribution,matsen1996unifying}, computer simulations \cite{gavrilov2013phase} and experiments \cite{khandpur1995polyisoprene}.

However, phase behavior of a single $(AB)_k$ multiblock copolymer chain placed in selective solvent is less clear. A number of authors observed microphase separation in a globule formed by a single $(AB)_k$ multiblock copolymer chain in a poor solvent. If A-type and B-type beads tended to segregate, microphase separation occurred leading to formation of unusual structures due to finite size of the globule \cite{theodorakis2011microphase, parsons2007single,theodorakis2011phase,ivanov1999computer}. A multiblock copolymer chain placed in a selective solvent, which is poor for B-type beads and good for A-type beads, was predicted to form either a swollen chain of molecular micelles or a single micelle depending on the length of a chain and its composition \cite{halperin1991collapse}. Results of computer simulations supported this prediction \cite{pham2010collapse,hugouvieux2009amphiphilic,lewandowski2008protein,rissanou2014collapse,ulianov2016active,woloszczuk2008alternating,wang2014coil}. Authors also observed formation of layered and tubular structures \cite{hugouvieux2009amphiphilic}, as well as dynamic switching between swollen chain of micelles and a single micelle \cite{rissanou2014collapse}.

However, the aforementioned simulation works \cite{pham2010collapse,hugouvieux2009amphiphilic,lewandowski2008protein,rissanou2014collapse,ulianov2016active,woloszczuk2008alternating,wang2014coil} had several limitations. First, either length of a block or the number of blocks in a multiblock copolymer chain was limited. Typical length of a block in the studied chains rarely exceeded 10 monomer unis (beads) \cite{pham2010collapse,hugouvieux2009amphiphilic,lewandowski2008protein,woloszczuk2008alternating,wang2014coil}. Authors of ref. \cite{rissanou2014collapse} studied chains with long blocks, but there were only 5 blocks in a chain. These limitations were overcome in the work \cite{ulianov2016active}, in which authors performed simulations of a multiblock copolymer chain consisting of many long blocks. However, the effect of composition of the chain, length of a block and strength of interactions on the chain conformation was not studied in this work. Second, the authors of refs. \cite{pham2010collapse,hugouvieux2009amphiphilic,lewandowski2008protein,rissanou2014collapse,ulianov2016active,woloszczuk2008alternating,wang2014coil,halperin1991collapse} studied the case of the so-called amphiphilic multiblock copolymers, in which attraction between A-type and B-type beads is either absent or significantly weaker than attraction between B-type beads. It is still unknown how strength of attraction between A-type and B-type beads affects the structure of a single multiblock copolymer chain.

On the other hand, a single multiblock copolymer chain with different types of interactions between beads is a promising model of chromatin organization in various organisms \cite{ulianov2016active,barbieri2012complexity,jost2014modeling}. We have proposed to model interactions between nucleosomes as reversible bonds in our previous works \cite{ulianov2016active,petrov2020kinetic}. In this study, we introduced a modified version of this model. It is known that lysine 16 in histone H4 (histone H4 "tail") can form a complex with acidic region on H2A-H2B histone dimer (the so-called "acidic patch") \cite{histoneinteractions,shahbazian2007functions}. We may treat this interaction as a reversible bond between two beads representing a nucleosome \cite{ulianov2016active,petrov2020kinetic}. Each nucleosome has two histone H4 tails and two acidic regions on the H2A-H2B histone dimer. It is also known that histone H4 tails are mostly acetylated in active chromatin \cite{shahbazian2007functions}. In addition, the energy gain of forming a complex with the acidic patch is much smaller for an acetylated histone H4 tail than for a non-acetylated one. Thus, affinity of acidic patches in nucleosomes in the actively transcribed regions will be high to the histone H4 tails in inactive chromatin. In some sense, nucleosomes from actively transcribed regions may act as a surfactant for inactive chromatin. We suggest a simple and robust treatment of these complex interactions between active (A) and inactive (B) regions of chromatin. An "inactive" nucleosome may form two reversible bonds with any nucleosome. However, an "activated" (via acetylation) nucleosome may form only one reversible bond and only with an "inactive" nucleosome.

In this work, we investigated the behavior of a single multiblock copolymer chain with various interactions between A and B blocks. We assessed how attraction between A-type beads and B-type beads affected the structure of the chain. If this interaction was purely repulsive, we observed formation of a chain of intramolecular micelles. However, an unusual compact structure was formed by the chain, which modeled interactions between nucleosomes by formation of reversible bonds between A-type and B-type beads. This structure had enrichment of contact density at the boundaries of A and B blocks, opposite to the situation observed in a chain of micelles, in which contact enrichment occurred inside the B blocks. We investigated the structure of such unusual state and described how distribution of contact density depended on the composition of a chain.

\section{Methods}
We studied a single flexible $(AB)_k$ multiblock copolymer chain of the length $N=10^4$ in an implicit solvent. The repeating unit of the copolymer consisted of one A and one B block (i.e. the copolymer sequence was $(A)_{nA}(B)_{nB}(A)_{nA}(B)_{nB}...$, where $n_A$ and $n_B$ were the lengths of A and B blocks, respectively). The total length of such repeating unit $(A)_{nA}(B)_{nB}$ was set to $n=n_A+n_B=400$. We characterized the copolymer composition by the fraction of B-type beads in the chain: $f = n_B/n$. We varied the value of $f$ from $f=0.2$ to $f=0.5$. Therefore, the lengths of the B-blocks varied from 80 to 200.

LAMMPS package was used in our work to perform Brownian dynamics simulations; the 12-6 Lennard-Jones potential (LJ) with the following parameters was applied: $\sigma=1.0$, $\epsilon=0.8$. To simulate purely repulsive interactions (i.e. good solvent conditions), we set the cutoff radius of the LJ potential to $R_\text{cut}=1.12$. To model poor solvent conditions, the cutoff radius was increased to $R_\text{cut}=2.00$, as, according to our preliminary simulations, such value was sufficient to observe the coil-to-globule transition of a homopolymer chain of the length $N=10^4$. In what follows, we denote the cutoff radius of LJ potential between beads of type X and Y ($X=A,B$, $Y=A,B$) as $R_\text{cut}^{XY}$.

Periodic boundary conditions were applied in our simulations, and the side of the cubic simulation box was equal to $350$. This choice ensured that a polymer chain of the length $N=10^4$ does not affect its own conformation in a good solvent.

We used the harmonic potential $U=K(r-r_0)^2$ to simulate the bonded interactions; the following parameters were used: $K=5.0$ and $b_0=0.5$. Under these conditions the chain was phantom, i.e. the bonds could easily cross each other. The same parameters were applied to simulate the backbone and the pairwise reversible bonds (see below).

In order to simulate the presence of reversible (dynamic) bonds between nucleosomes, we used the standard stochastic procedure for the creation and removal of bonds implemented in LAMMPS. These bonds were created in addition to the existing bonds in the chain backbone. Such bonds were created and broken every $N_\text{stp}=200$ MD steps. The probability of bond formation was fixed and equal to $1$, and the probability of breaking a bond was equal to $0.1$. The distance within which reversible bonds could be formed was set to $R_\text{max}^\text{create}=1.30$. This value was chosen so that the average lengths of the forming and breaking bonds were approximately equal: $<r_\text{form}>=1.1599$, $<r_\text{break}>=1.1601$. In addition, according to our preliminary simulations, these parameters led to the collapse of a homopolymer chain placed in athermal solvent if the monomer units could form maximum two reversible bonds with each other.

In order to characterise the conformation of a polymer chain in different conditions, we calculated the dependencies of the average spatial distance between two monomer units on the distance between them along the chain, $R(s)$. We also determined the dependencies of the contact probability of the monomer units on the distance along the chain, $P(s)$, and the contact maps.

In order to generate the initial chain conformations, we performed equilibration of a chain without reversible bonds in good solvent for $t_1=1.5\times 10^8$ MD steps. After this equilibration, we performed simulations for $t_2=1.5\times 10^8$ time steps under desired system conditions, and the last $5\times 10^7$ steps of that run were used to average $R(s)$ and $P(s)$ (11 conformations were used for averaging in total). To average the contact maps, we also performed 10 additional runs from different initial chain conformations. Therefore, the total number of chain conformations used for calculation of the contact maps was equal to $11\times 11=121$.

To obtain the contact probabilities $P(s)$, we calculated $K(s)$: the total number of bead pairs separated by $s$ beads along the chain and located closer than $r_\text{c}=1.5$ to each other in space. The contact probability was calculated as $P(s) = K(s)/(N-s)$. A similar methodology was already used to study collapse of a homopolymer globule \cite{chertovich2014crumpled}.

To analyze the heterogeneous distribution of contacts, we calculated $P(s)$ dependencies separately for the contacts between A-type beads ($P_\text{AA}(s)$), A-type and B-type beads ($P_\text{AB}(s)$), and B-type beads ($P_\text{BB}(s)$). To obtain a $P_\text{XY}(s)$ dependency ($X=A,B$, $Y=A,B$), we calculated the $K(s)$ value only for the beads of the types X and Y $K_\text{XY}(s)$, and then divided $K_\text{XY}(s)$ by the total number of bead pairs of type X and Y separated by $s$ beads along the chain. For brevity, further we denote contacts or interactions between beads of type $X$ and beads of type $Y$ as XY contacts or interactions, respectively.

Contact maps are a visual tool to show how frequently different parts of the chain are in contact with each other. On its $i\times j$ position, a contact map contained the contact probability of the $i$th and $j$th monomer units. Since the simulated chain was rather long, we performed coarsening of the contact map for better visual appearance, so the resulting contact maps had the size of $0.1N \times 0.1N$. Further we denote the element of the coarsened contact map as $p(i,j)$. In addition, we performed the "sliding window" averaging procedure to obtain the averaged probability of contact between beads lying within five repeating units. To calculate the element of the contact map after the "sliding window" $p_{sliding}(i,j)$, $|i-j|<200$, we followed the following procedure. First, we added up the $p(i+kn, j+kn)$ values for all possible values of integer $k$, $k\in [1; 1 + N/n - 5]$. Second, we obtained $p_{sliding}(i,j)$ by dividing this sum by $1 + N/n - 5$.

\section{Results}
We have investigated the behavior of a multiblock copolymer chain with different interactions between beads. We describe structure of a multiblock copolymer chain with beads interacting only via LJ potential in the section III.A. The behavior of a chain placed in athermal solvent (all LJ interactions are purely repulsive) with reversible bonds forming between beads is described in the section III.B.

\subsection{Volume Interactions}
In this section, we describe the behavior of chains, in which either only BB, or BB and AB interactions are attractive. AA interactions are purely repulsive.

\begin{figure}[htb]
\centering
    \begin{subfigure}{0.49\textwidth}
	\includegraphics[width=\linewidth,height=\textheight,keepaspectratio]{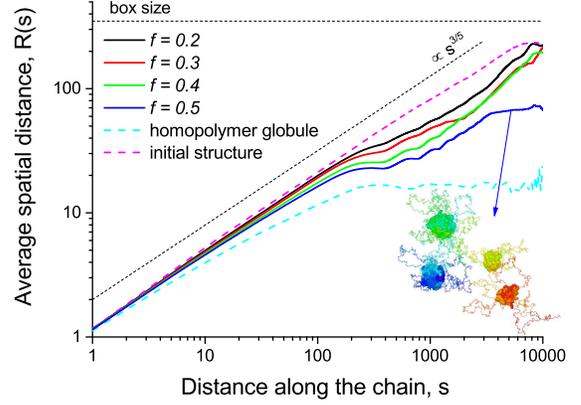}
	\caption{}
	\end{subfigure}
	\begin{subfigure}{0.49\textwidth}
	\includegraphics[width=\linewidth,height=\textheight,keepaspectratio]{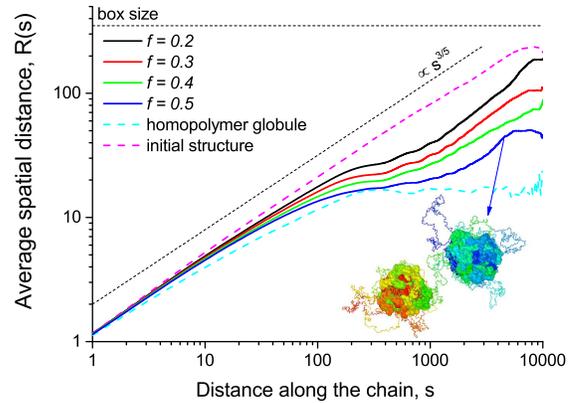}
	\caption{}
	\end{subfigure}
	\caption{$R(s)$ dependencies in multiblock copolymer chains with different fraction of B-type beads $f$, attraction is realized via LJ potential. Snapshots of structures formed by a chain with $f=0.5$ are shown on each figure. Thin lines represent B-type beads, spheres represent A-type beads. The beads are colored according to their position along the chain. (a) Only BB interaction is attractive ($R_\text{cut}^{BB}=2.0$), other interactions are purely repulsive. (b) AA interactions are purely repulsive, AB and BB interactions are attractive (i.e. $R_\text{cut}^{AB}=R_\text{cut}^{BB}=2.0$).}
    \label{rsvol}
\end{figure}

\begin{figure}[htb]
\centering
    \begin{subfigure}{0.49\textwidth}
	\includegraphics[width=\linewidth,height=\textheight,keepaspectratio]{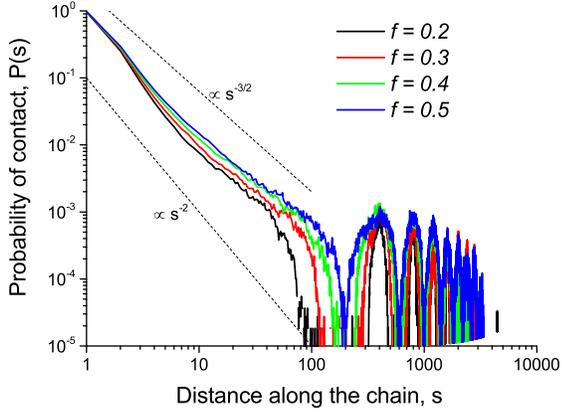}
	\caption{}
	\end{subfigure}
	\begin{subfigure}{0.49\textwidth}
	\includegraphics[width=\linewidth,height=\textheight,keepaspectratio]{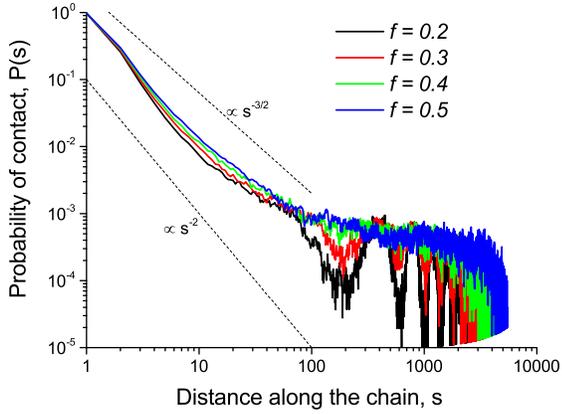}
	\caption{}
	\end{subfigure}
	\caption{$P(s)$ dependencies in multiblock copolymer chains with different fraction of B-type beads $f$, attraction is realized via LJ potential. (a) Only BB interaction is attractive ($R_\text{cut}^{BB}=2.0$), other interactions are purely repulsive. (b) AA interactions are purely repulsive, AB and BB interactions are attractive (i.e. $R_\text{cut}^{AB}=R_\text{cut}^{BB}=2.0$).}
    \label{psvol}
\end{figure}

First, we studied the behavior of a chain, in which only BB interaction was attractive. Other interactions were purely repulsive. To model such interactions, we set the cutoff radius of LJ potential acting between B-type beads to $R_\text{cut}^{BB}=2.0$. Cutoff radii of LJ potentials acting between other bead pairs were set to $R_\text{cut}^{AB}=R_\text{cut}^{AA}=1.12$. This case was well studied previously by theory and computer simulations and is the simplest model of a multiblock copolymer chain in highly selective solvent \cite{pham2010collapse,hugouvieux2009amphiphilic,lewandowski2008protein,rissanou2014collapse,ulianov2016active,woloszczuk2008alternating,wang2014coil,halperin1991collapse}. We observed formation of molecular micelles along the chain as predicted by theory \cite{halperin1991collapse}. The $R(s)$ dependencies of such conformations had a characteristic step-like shape, which supported visual observation of formation of molecular micelles along the chain (Fig. \ref{rsvol}a). The $P(s)$ dependencies had oscillations demonstrating local aggregation of blocks into micelles (Fig. \ref{psvol}a). It is worth mentioning that probability of contact was equal almost to zero starting from a certain large $s$ (Fig. \ref{psvol}a). This finding is not surprising, since the chain forms a self-avoiding walk of molecular micelles \cite{halperin1991collapse}.

\begin{figure}[htbp]
    \centering
	\includegraphics[width=\linewidth,keepaspectratio]{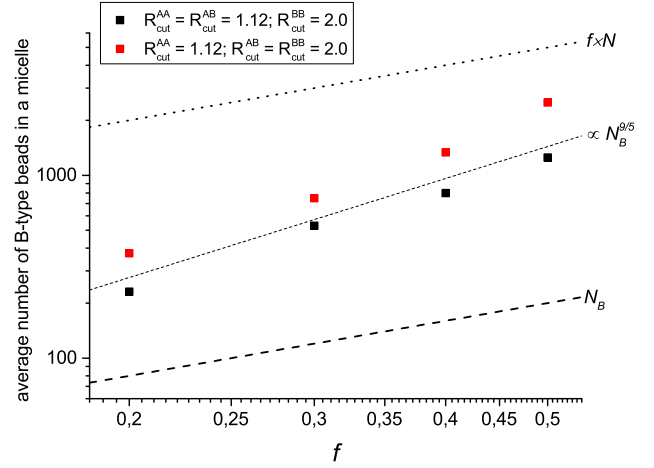}
	\caption{The dependency of the average number of B-type beads in a micelle on the fraction of B-type beads in a chain $f$. Black dots: only BB interaction is attractive ($R_\text{cut}^{BB}=2.0$), other interactions are purely repulsive. Red dots: AA interactions are purely repulsive, AB and BB interactions are attractive (i.e. $R_\text{cut}^{AB}=R_\text{cut}^{BB}=2.0$). Dashed and dotted lines represent the minimal and maximal possible values for the number of B-type beads in a micelle, respectively.}
    \label{blockvol}
\end{figure}

Halperin predicted how the average number of B-type beads in a micelle $N_\text{Bmic}$ scales with the number of B-type beads in a block: $N_\text{Bmic}\propto N_B^{9/5}$ \cite{halperin1991collapse}. We have observed an excellent agreement of simulation results with this prediction (Fig. \ref{blockvol}).

Second, we studied the structure of a multiblock copolymer chain with strong attraction between A-type and B-type beads ($R_\text{cut}^{AB}=R_\text{cut}^{BB}=2.0$). Interactions between A-type beads were purely repulsive. We observed that the chain still resembled a chain of micelles (Fig. \ref{rsvol}b, \ref{psvol}b), but the number of B-type beads comprising one micelle was larger than in the previous case (Fig. \ref{blockvol}). This indicates that a certain portion of A-type beads acted as a "glue", "sticking" the B-type beads together. It is worth mentioning that $N_\text{Bmic}$ scaled similarly with $f$ as in the system without AB attraction (Fig. \ref{blockvol}). Therefore, the number of B-type beads comprising one micelle increased by a constant factor independent of $f$ after AB attraction had been switched on.

\subsection{Reversible Bonds}
In this section, we describe structure of a multiblock copolymer chain placed in athermal solvent (all volume interactions are purely repulsive, $R_\text{cut}^{AA} = R_\text{cut}^{AB}=R_\text{cut}^{AA}=1.12$). Reversible bonds may form between beads. Average lifetime of a bond is equal to $\tau=2\times 10^3$ MD steps, probability of bond formation is equal to unity.

\begin{figure}[htb]
\centering
    \begin{subfigure}{0.49\textwidth}
	\includegraphics[width=\linewidth,height=\textheight,keepaspectratio]{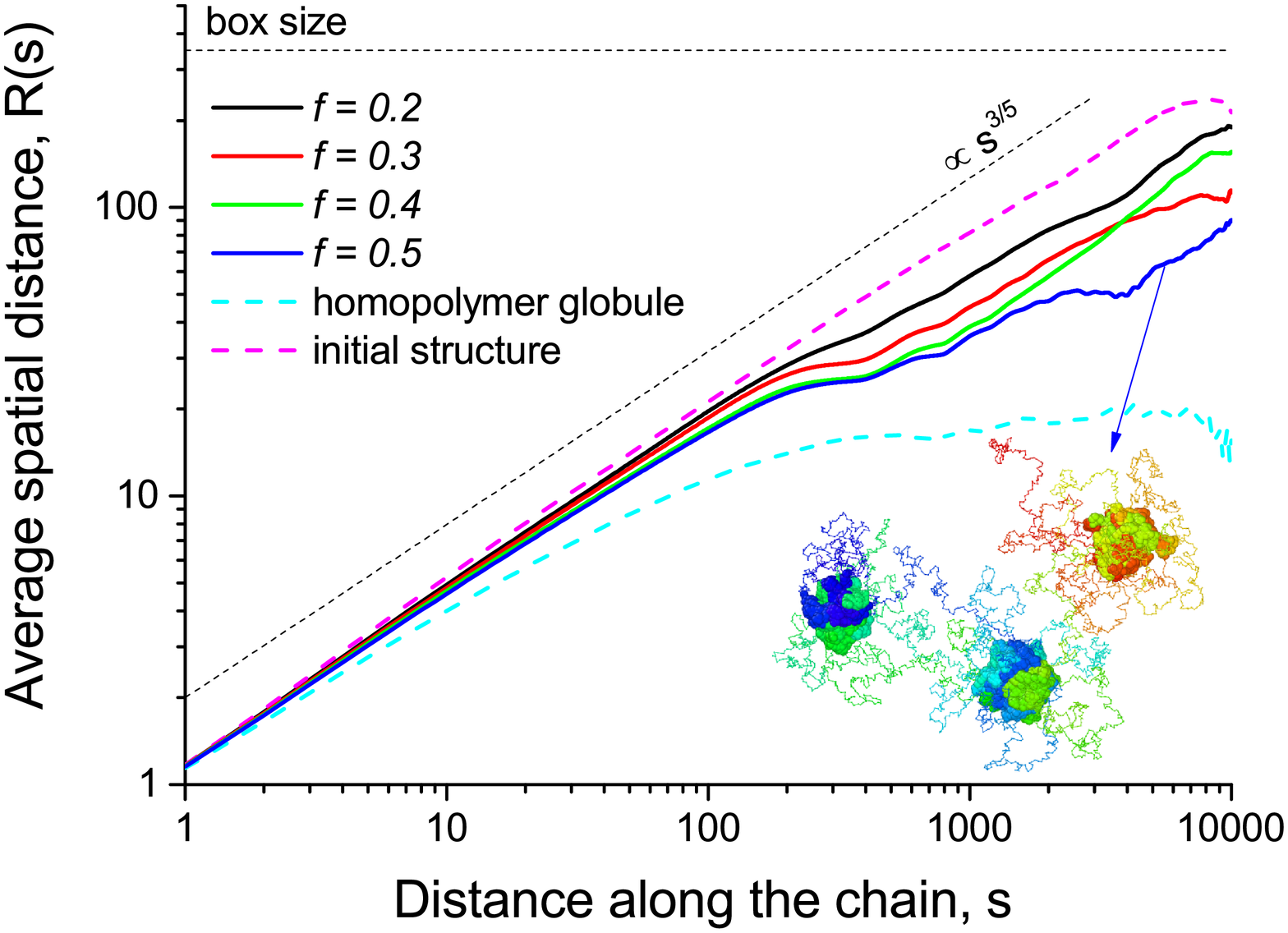}
	\caption{}
	\end{subfigure}
	\begin{subfigure}{0.49\textwidth}
	\includegraphics[width=\linewidth,height=\textheight,keepaspectratio]{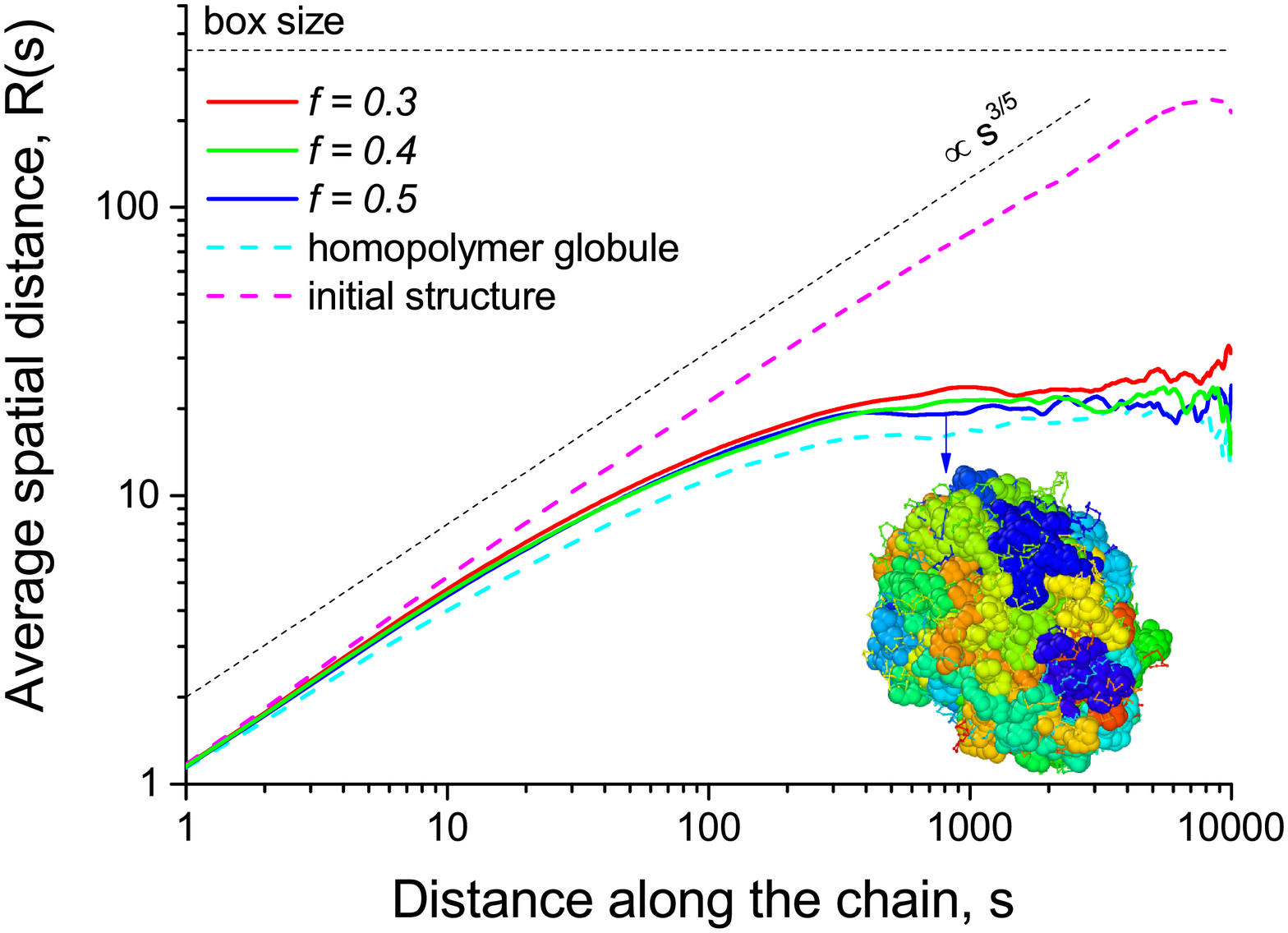}
	\caption{}
	\end{subfigure}
	\caption{$R(s)$ dependencies in multiblock copolymer chains with different fraction of B-type beads $f$. Snapshots of structures formed by a chain with $f=0.5$ are shown on each figure. Thin lines represent B-type beads, spheres represent A-type beads. The beads are colored according to their position along the chain. (a) All interactions were purely repulsive with one exception: B-type beads could form maximum two reversible bonds with each other. (b) Interactions between A-type beads were purely repulsive. A-type beads could form maximum one bond with a B-type bead, B-type beads could form maximum two bonds either with A-type or B-type beads.}
    \label{rssat}
\end{figure}

\begin{figure}[htb]
\centering
    \begin{subfigure}{0.49\textwidth}
	\includegraphics[width=\linewidth,height=\textheight,keepaspectratio]{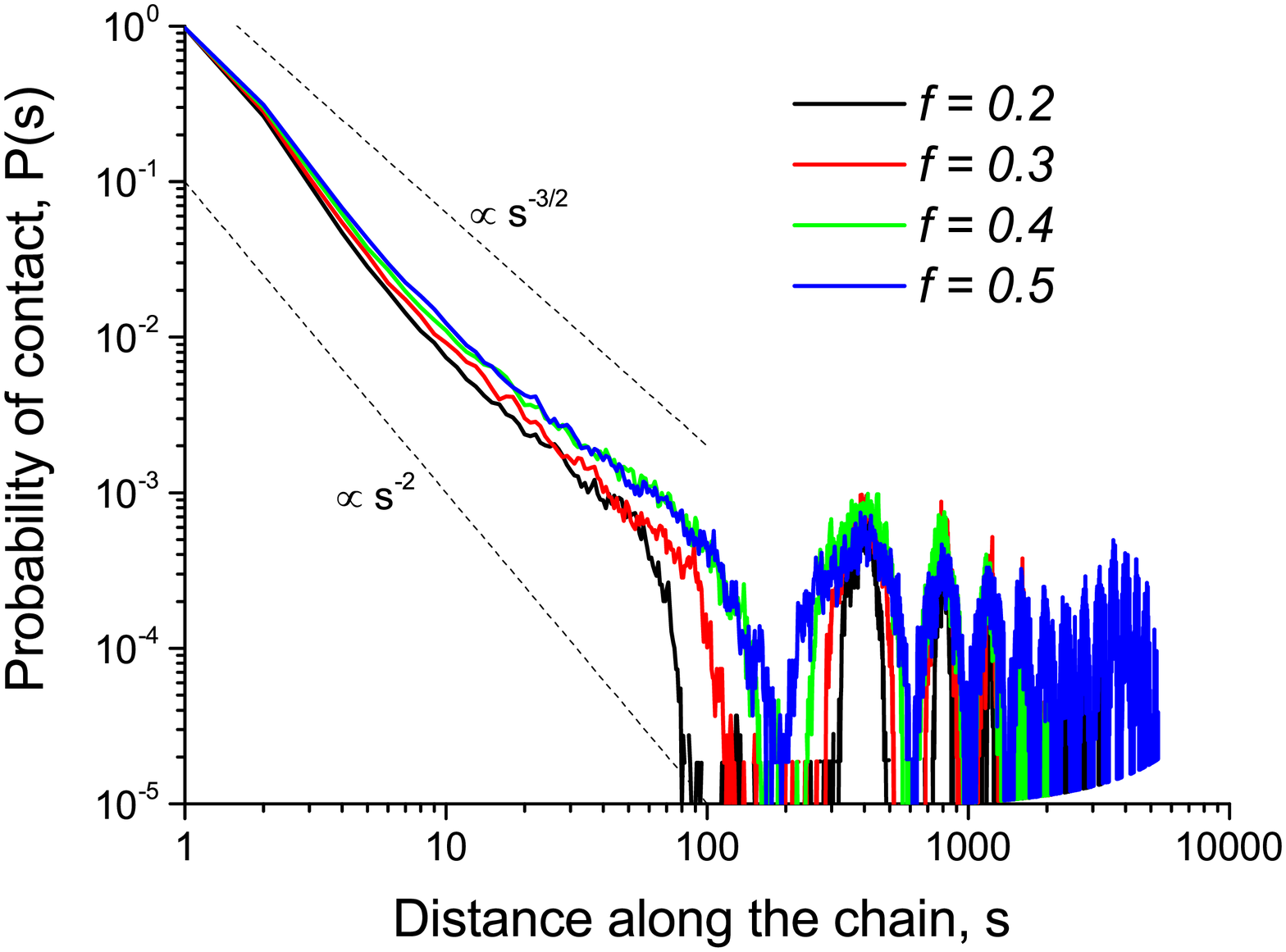}
	\caption{}
	\end{subfigure}
	\begin{subfigure}{0.49\textwidth}
	\includegraphics[width=\linewidth,height=\textheight,keepaspectratio]{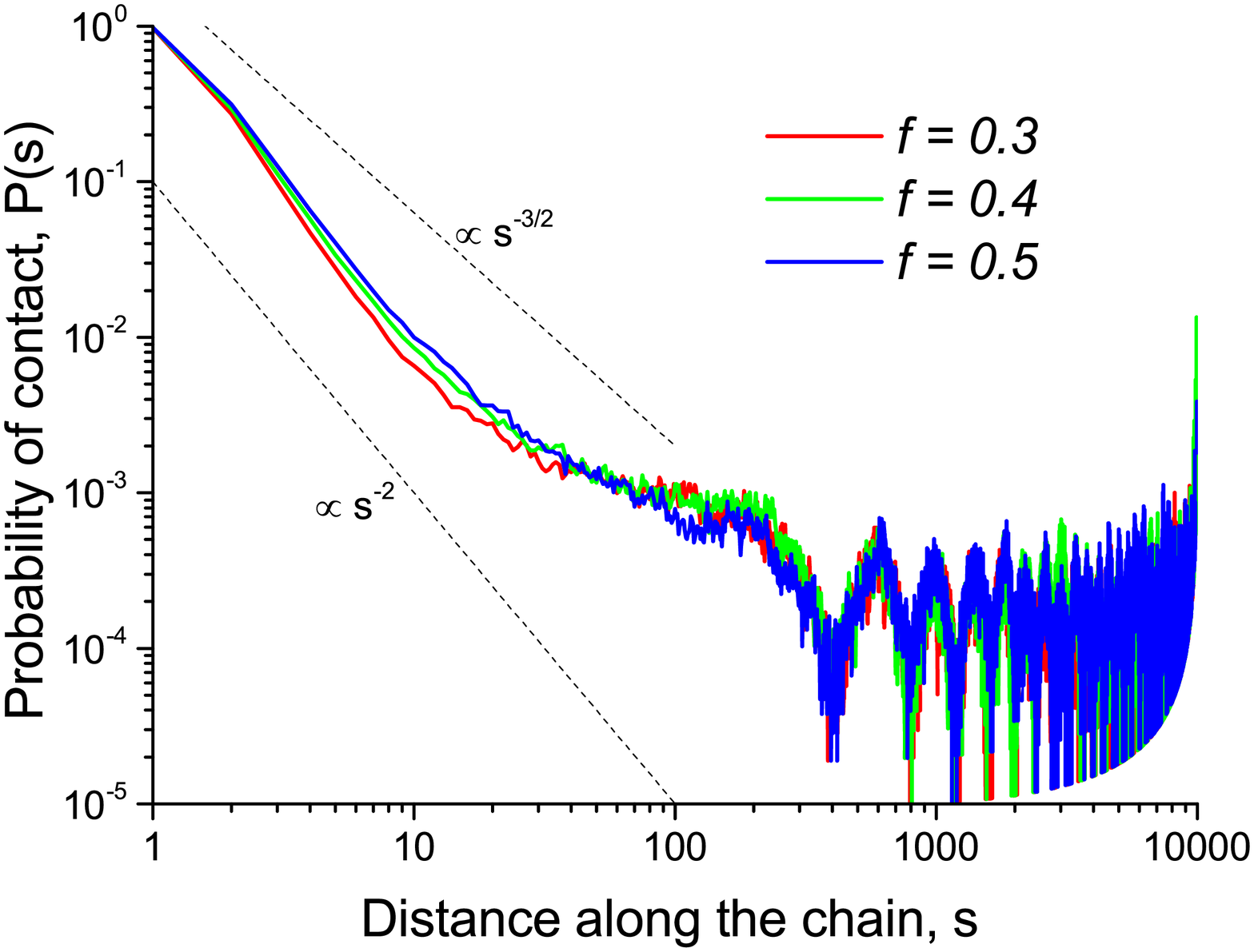}
	\caption{}
	\end{subfigure}
	\caption{$P(s)$ dependencies in multiblock copolymer chains with different fraction of B-type beads $f$. (a) All interactions were purely repulsive with one exception: B-type beads could form maximum two reversible bonds with each other. (b) Interactions between A-type beads were purely repulsive. A-type beads could form maximum one bond with a B-type bead, B-type beads could form maximum two bonds either with A-type or B-type beads.}
    \label{pssat}
\end{figure}

\begin{figure}[htbp]
    \centering
	\includegraphics[width=\linewidth,keepaspectratio]{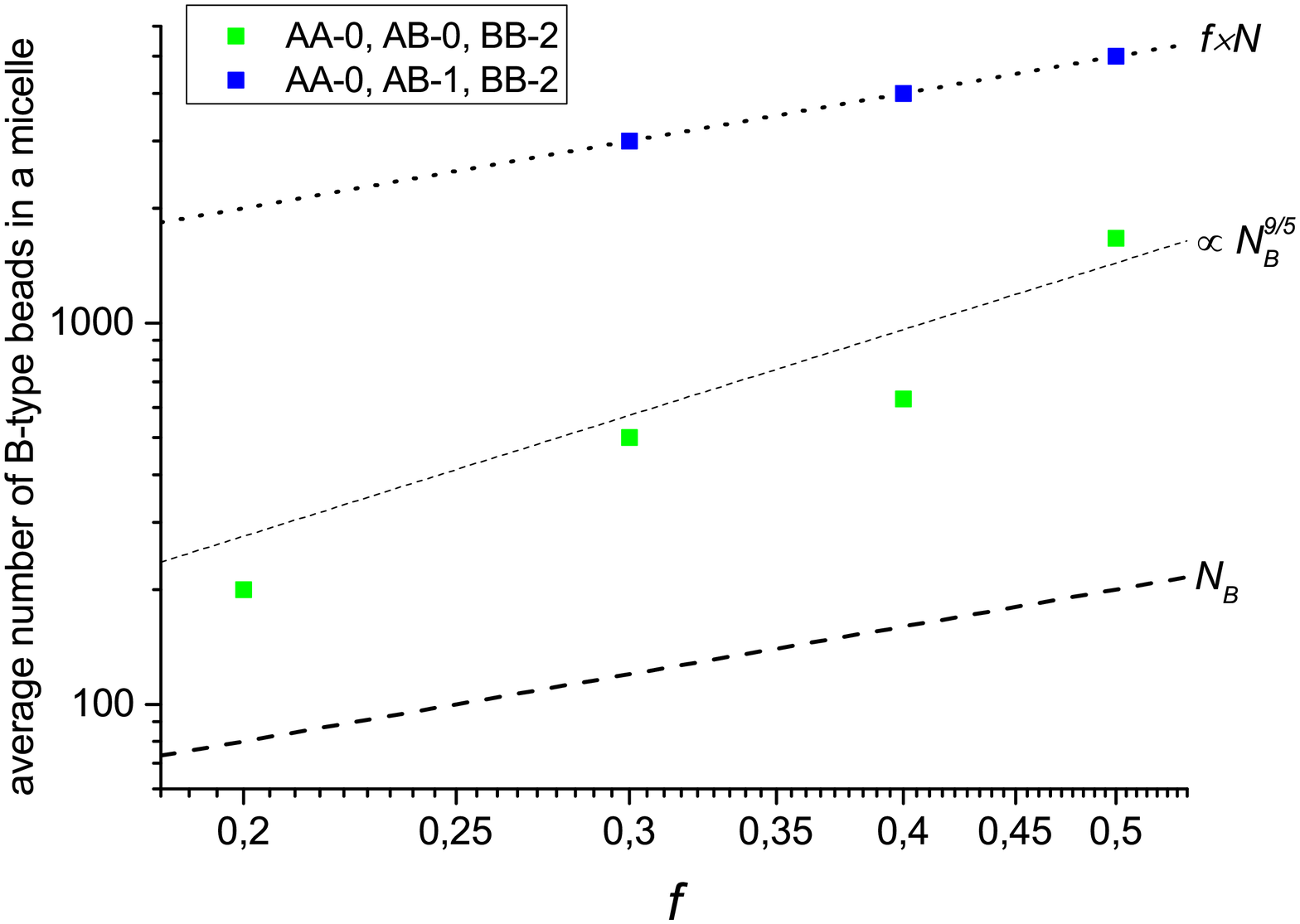}
	\caption{The dependency of the average number of B-type beads in a micelle on the fraction of B-type beads in a chain $f$. Green dots: all interactions were purely repulsive with one exception: B-type beads could form maximum two reversible bonds with each other. Blue dots: interactions between A-type beads were purely repulsive. A-type beads could form maximum one reversible with a B-type bead. B-type beads could form maximum two reversible bonds either with A-type or B-type beads. Dashed and dotted lines represent the minimal and maximal possible values for the number of B-type beads in a micelle, respectively.}
    \label{blocksat}
\end{figure}

The first studied system was a multiblock copolymer chain, in which B-type beads could form maximum two reversible bonds only with each other. Other interactions were purely repulsive. In this case, the chain formed a string of micelles (Fig. \ref{rssat}a, \ref{pssat}a). We have observed qualitatively similar structures as in the chain, in which attraction between B-type beads was realized via LJ potential. The only difference we observed is a slight mixing of two micelles in the chain with $f=0.5$ (Fig. \ref{rssat}a). The average number of B-type beads in a micelle scaled approximately as predicted by theory \cite{halperin1991collapse} (Fig. \ref{blocksat}).

The second studied case was a multiblock copolymer chain, in which an A-type bead could form maximum one reversible bond with a B-type bead. A B-type bead could form maximum two bonds with either A-type or B-type beads. Surprisingly, the chain behaved qualitatively different from the previous cases. First of all, the chain did not form a string of intramolecular micelles with a well-defined core constituted by the B-type beads and a corona consisting of A-type beads. Instead, we observed formation of a compact structure, since $R(s)$ dependencies reached plateau for large $s$ (Fig. \ref{rssat}b). In addition, B-type beads formed a single cluster for all values of $f$ (Fig. \ref{blocksat}). We did not study the case $f=0.2$, since formation of a unified collapsed structure was hampered due to insufficient number of B-type beads and therefore strong fluctuations of the structure.

To study the collapsed structure in more detail, we built the $P(s)$ dependencies (Fig. \ref{pssat}b). Contact probability did not turn to zero for all $s$ (Fig. \ref{pssat}b) thus suggesting the globular structure of the chain in accord with Fig. \ref{rssat}b. Notably, we observed periodic behavior of the dependency for all values of $f$. Therefore, the globular structure formed in this case had internal heterogeneous distribution of contact density.

\begin{figure}[H]
\centering
    \begin{subfigure}{0.49\textwidth}
	\includegraphics[width=\linewidth,height=\textheight,keepaspectratio]{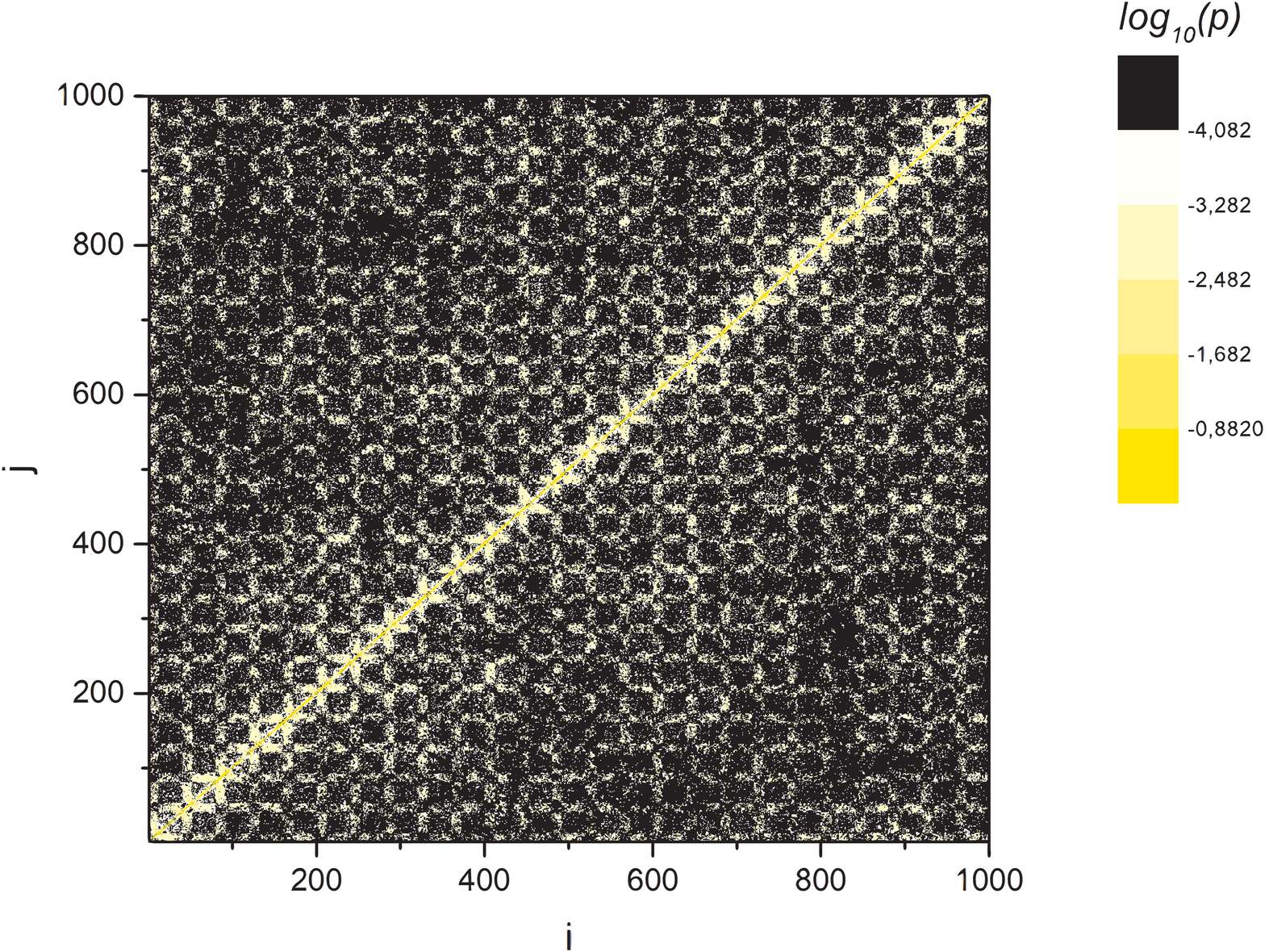}
	\caption{}
	\end{subfigure}
	\begin{subfigure}{0.49\textwidth}
	\includegraphics[width=\linewidth,height=\textheight,keepaspectratio]{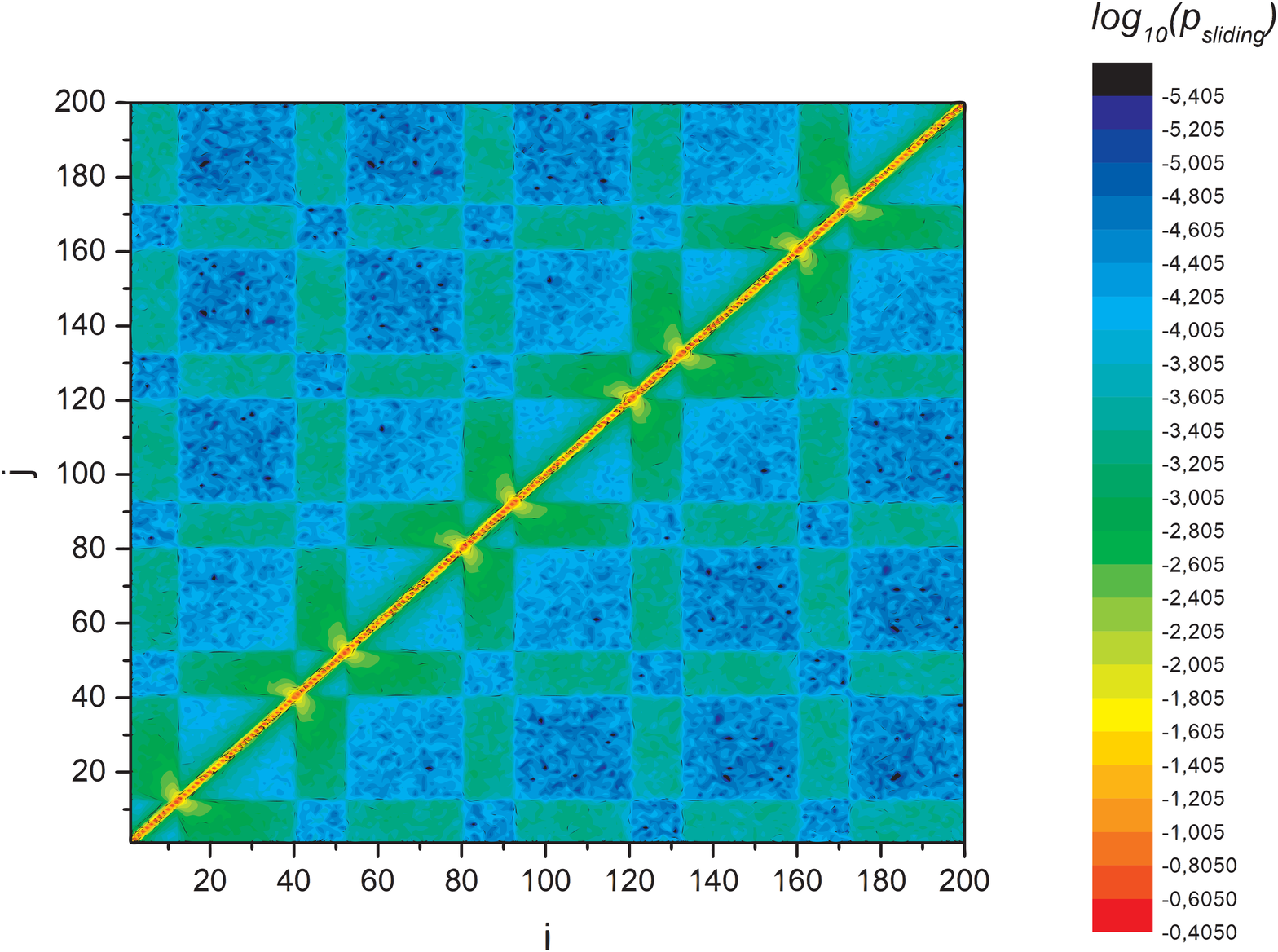}
	\caption{}
	\end{subfigure}
	\begin{subfigure}{0.49\textwidth}
	\includegraphics[width=\linewidth,height=\textheight,keepaspectratio]{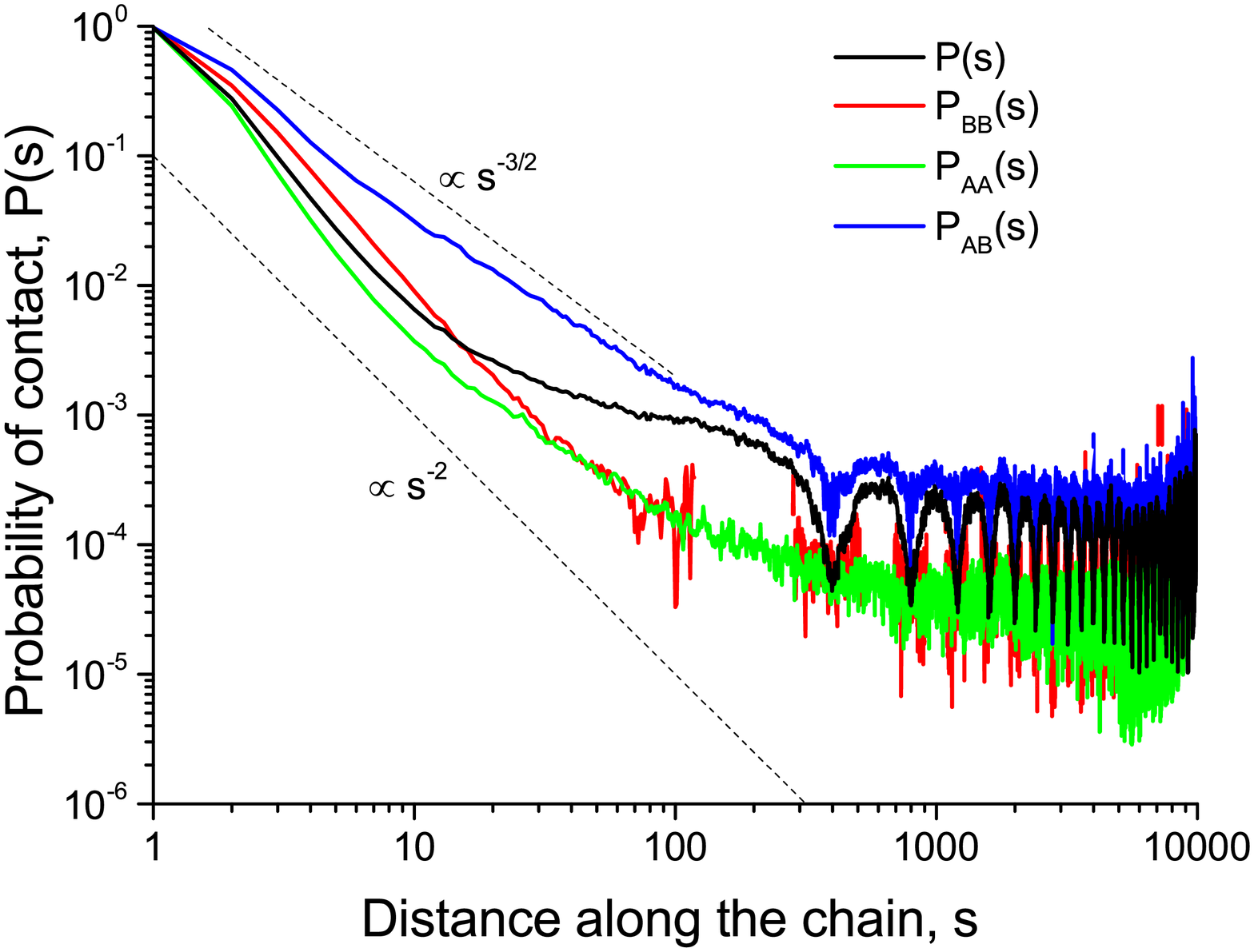}
	\caption{}
	\end{subfigure}
	\caption{Coarsened contact map (a), contact map after "sliding window" averaging (b), and the dependencies of contact probability on the distance along the chain (c) for a multiblock copolymer chain with reversible bonds, $f=0.3$. Interactions between A-type beads are purely repulsive. A-type beads could form maximum one bond with a B-type bead, B-type beads could form maximum two bonds either with A-type or B-type beads. The data was averaged over 11 initial conformations.}
    \label{pssat_2}
\end{figure}

\begin{figure}[H]
\centering
    \begin{subfigure}{0.49\textwidth}
	\includegraphics[width=\linewidth,height=\textheight,keepaspectratio]{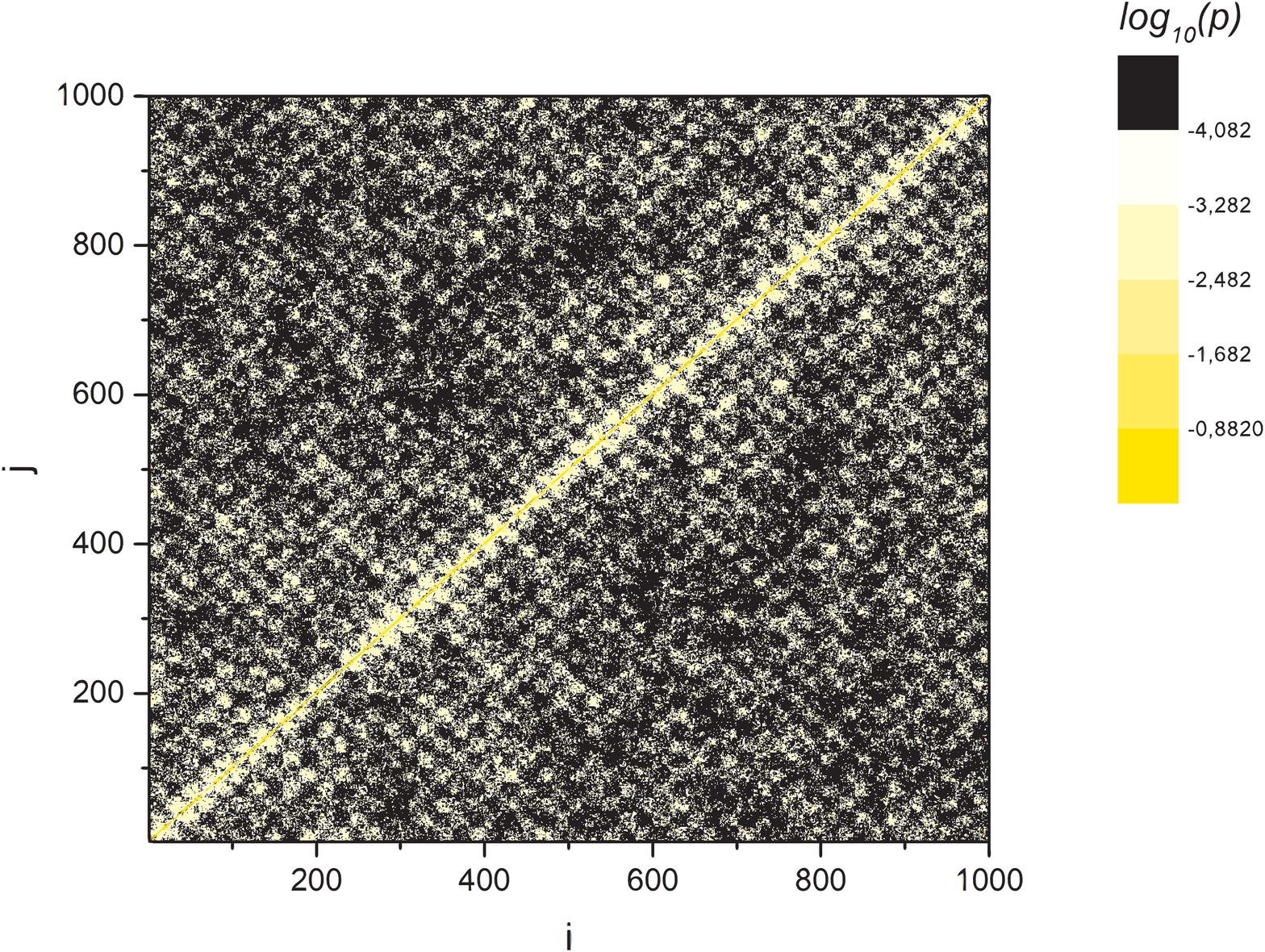}
	\caption{}
	\end{subfigure}
	\begin{subfigure}{0.49\textwidth}
	\includegraphics[width=\linewidth,height=\textheight,keepaspectratio]{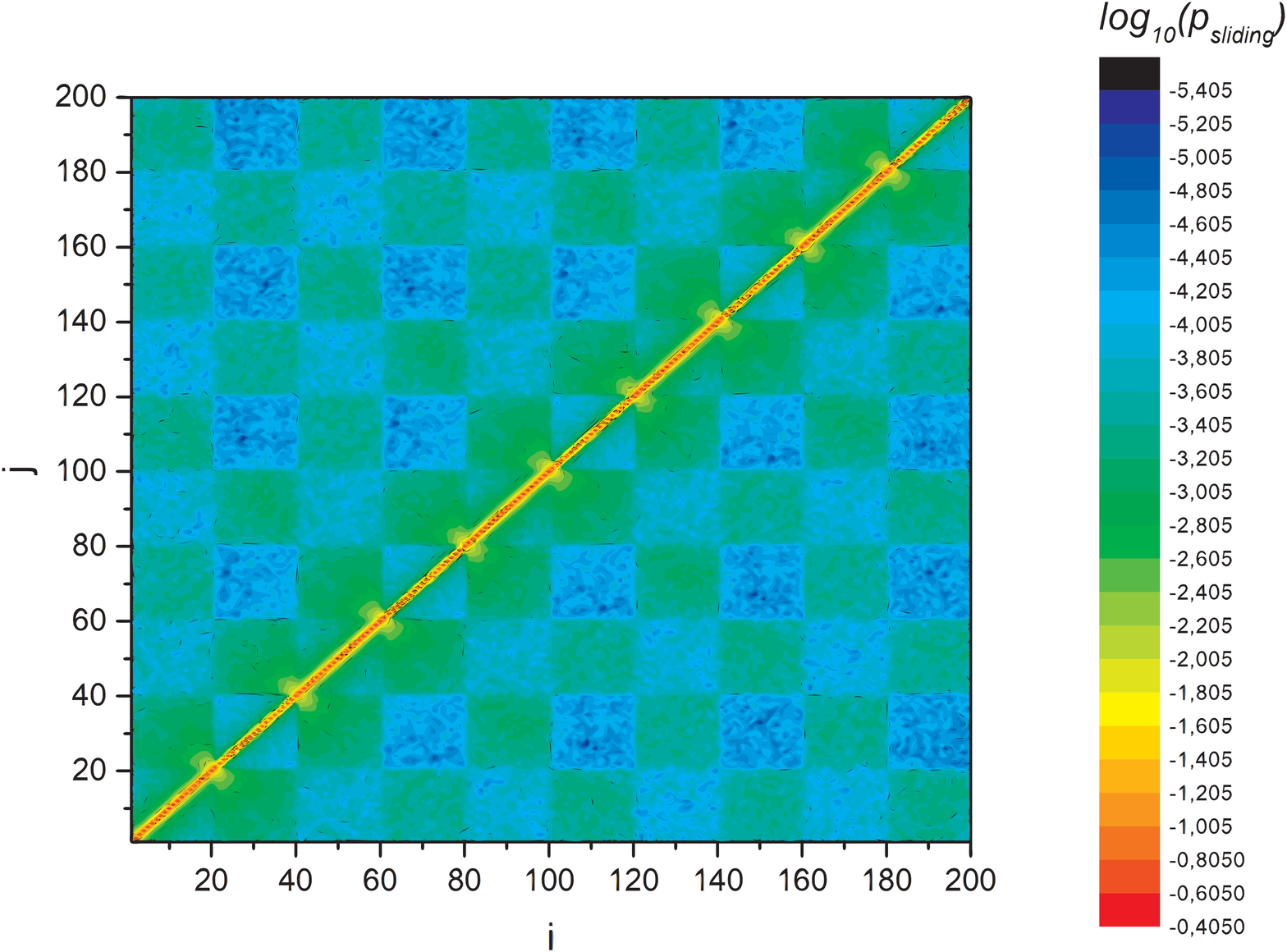}
	\caption{}
	\end{subfigure}
	\begin{subfigure}{0.49\textwidth}
	\includegraphics[width=\linewidth,height=\textheight,keepaspectratio]{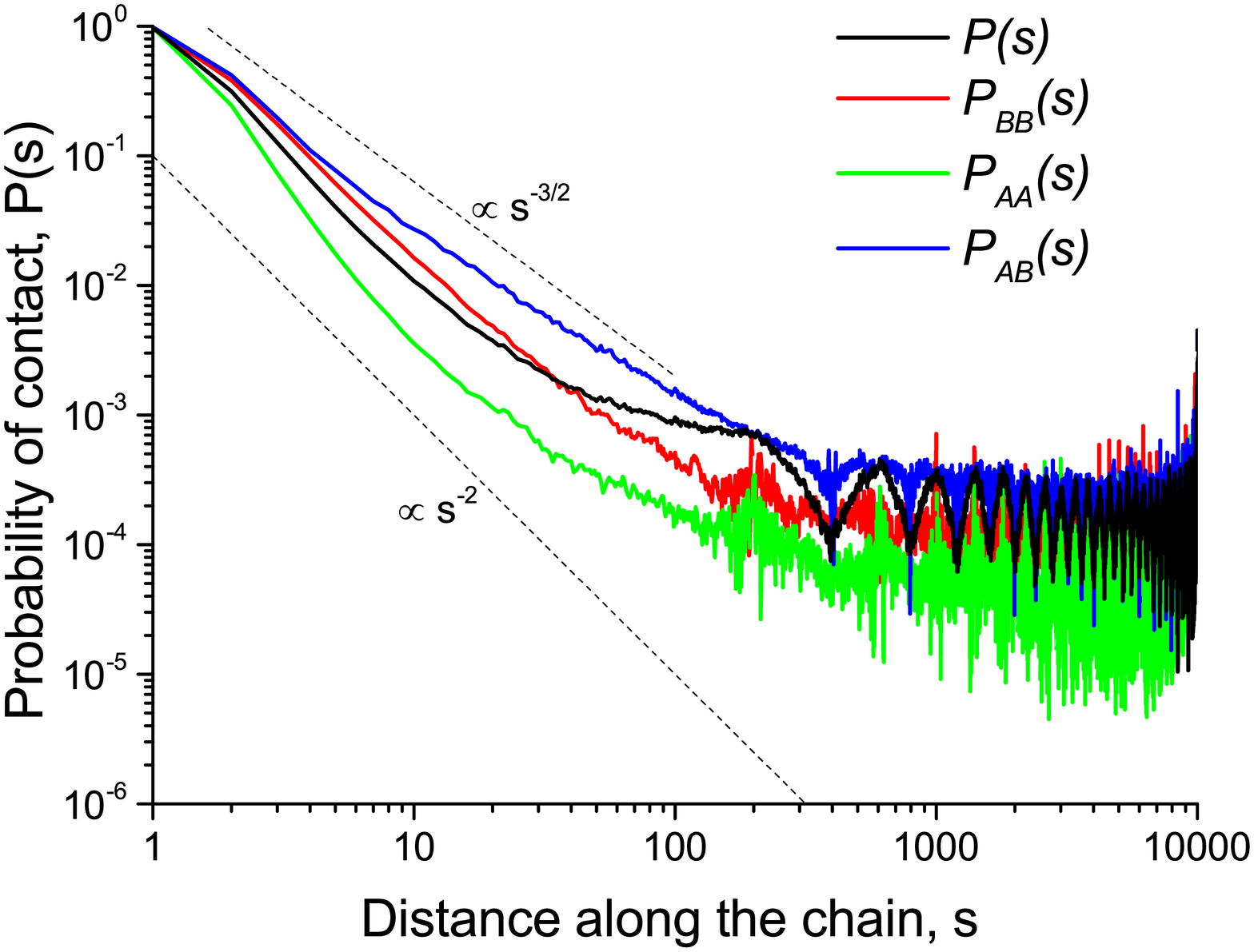}
	\caption{}
	\end{subfigure}
	\caption{Coarsened contact map (a), contact map after "sliding window" averaging (b), and the dependencies of contact probability on the distance along the chain (c) for a multiblock copolymer chain with reversible bonds, $f=0.5$. Interactions between A-type beads are purely repulsive. A-type beads could form maximum one bond with a B-type bead, B-type beads could form maximum two bonds either with A-type or B-type beads. The data was averaged over 11 initial conformations.}
    \label{pssat_3}
\end{figure}

We also built contact maps for the chains with $f=0.3$ and $f=0.5$ (Fig. \ref{pssat_2}a, \ref{pssat_3}a). Contact maps exhibited a unique checkerboard-like structure, demonstrating enrichment of contacts between A-type and B-type beads (yellow "stripes") and depletion of contacts between blocks containing beads of the same type (black "holes"). Contact maps after "sliding window" averaging procedure demonstrated this picture even more clearly (Fig. \ref{pssat_2}b, \ref{pssat_3}b). We also analyzed the dependencies of contact probability between beads of specific type $P_\text{XY}(s)$ ($X=A,B$, $Y=A,B$) as described in Methods (Fig. \ref{pssat_2}c, \ref{pssat_3}c). These dependencies demonstrated that AB contacts occurred much more frequently than AA or BB contacts. Moreover, the $P(s)$ dependencies had local minima at $s\approx q\times n$, where $q$ is an integer (Fig. \ref{pssat_2}c, \ref{pssat_3}c). Our data suggested that contact enrichment occurred at the boundaries of A and B blocks.

We also studied how this unusual structure depended on the fraction of B-type beads in a chain $f$. We observed that in the chain with $f=0.5$ the checkerboard-like pattern contained three types of "squares" on the contact map (Fig. \ref{pssat_3}b). AB contacts occurred with the highest probability (green squares), BB contacts had intermediate frequency of occurrence (cyan squares), and AA contacts had the lowest probability of occurrence (squares containing blue dots,  Fig. \ref{pssat_3}b). We did not observe such hierarchy of contact frequencies in the chain with $f=0.3$, in which probabilities of AA and BB contacts were almost equal on the large scale (Fig. \ref{pssat_2}c). This data suggests that we can govern distribution of contacts within the structure by altering the chain composition.

\section{Discussion}
In this work, we have studied structure of a single multiblock copolymer chain. We have shown that behavior of a chain becomes very nontrivial if B-type beads can form reversible bonds not only with B-type beads, but also with A-type beads. Such chain can also be treated as a model of chromatin organization, since reversible bonds can model interactions between nucleosomes in hetero- and euchromatin. Our data has demonstrated that the structure formed by such a multiblock copolymer chain has a unique internal distribution of contacts depending on the chain composition. Contact enrichment occurs at the boundaries of the blocks, and depletion of contact density is observed within the A and B blocks. Further research is needed to elucidate physical mechanisms of formation of such structures.

\section*{Acknowledgements}
We thank Pavel Kos for fruitful discussions and comments. The research is carried out using the equipment of the shared research facilities of HPC computing resources at Lomonosov Moscow State University. The reported study was funded by RFBR according to the research project \# 18-29-13041.

\bibliography{sample}
\end{document}